\documentclass[10pt,journal,compsoc]{IEEEtran}
\usepackage{amsmath,amsfonts}
\usepackage{algorithmic}
\usepackage{array}
\usepackage[caption=false,font=normalsize,labelfont=sf,textfont=sf]{subfig}
\usepackage{textcomp}
\usepackage{stfloats}
\usepackage{url}
\usepackage{verbatim}
\usepackage{graphicx}
\usepackage{cite}
\usepackage[backref]{hyperref}
\usepackage{multirow}
\usepackage{underscore}
\usepackage{color}
\usepackage{colortbl}
\usepackage{xcolor}
\usepackage[justification=centering]{caption}
\usepackage{enumitem}

\usepackage{setspace}

\usepackage{picins}

\usepackage{verbatim}
\usepackage{enumitem}
\usepackage{makecell}
\usepackage{hyperref}

\usepackage{booktabs}    
\usepackage{multirow}    
\usepackage{diagbox}
\usepackage{bbding}

\def\BibTeX{{\rm B\kern-.05em{\sc i\kern-.025em b}\kern-.08em
    T\kern-.1667em\lower.7ex\hbox{E}\kern-.125emX}}

\usepackage{balance}
\begin{document}
\title{A Chain of AI-based Solutions for Resolving FQNs and Fixing Syntax Errors in Partial Code}
\author{Qing~Huang,
        Jiahui~Zhu,
        Zhenchang~Xing,
        Huan~Jin,
        Changjing~Wang,
        Xiwei~Xu
\IEEEcompsocitemizethanks{\IEEEcompsocthanksitem Q. Huang, J. Zhu, C. Wang are with School of Computer Information Engineering, Jiangxi Normal University, China.\protect
\IEEEcompsocthanksitem H. Jin is with Jiangxi University of Technology’s School of Information Engineering, China.
\IEEEcompsocthanksitem Q. Huang and J. Zhu are co-first authors, C. Wang is the corresponding author (wcj@jxnu.edu.cn)
\IEEEcompsocthanksitem Z. Xing is with Australian National University, Australia.
\IEEEcompsocthanksitem X. Xu is with Data61, CSIRO.}
}

\IEEEtitleabstractindextext{
\begin{abstract}
API documentation, technical blogs and programming Q\&A sites contain numerous partial code that can be reused in programming tasks, but often these code are uncompilable due to unresolved names and syntax errors. 
To facilitate partial code reuse, we propose the Partial Code Reuse Chain (PCR-Chain) for resolving fully-qualified names (FQNs) and fixing last-mile syntax errors in partial code based on a giant large language model (LLM) like ChatGPT. Methodologically, PCR-Chain is backed up by the underlying global-level prompt architecture (which combines three design ideas: hierarchical task breakdown, prompt composition, and a mix of prompt-based AI and non-AI units) and the local-level prompt design. 
Technically, we propose PCR-Chain, which employs in-context learning rather than symbolic, costly training methods. Experimental results demonstrate that in dynamically-typed languages (Python), PCR-Chain outperforms current state-of-the-art (SOTA) 5\% accuracy like RING.
For statically-type languages (Java), our approach achieves high accuracy of 80.5\% in resolving both non-FQNs and last-mile syntax errors, surpassing SOTA methods (RING) that can only address last-mile syntax errors.
The correct execution of the unit, module, and PCR-Chain demonstrates the effectiveness of the prompt design, composition, and architecture and opens up possibilities for building software engineering tools based on LLMs, replacing traditional program analysis methods.
\end{abstract}
}

\maketitle
\IEEEdisplaynontitleabstractindextext
\IEEEpeerreviewmaketitle

\section{Introduction}

\begin{figure*}
    \centering
    \includegraphics[width=0.9\textwidth]{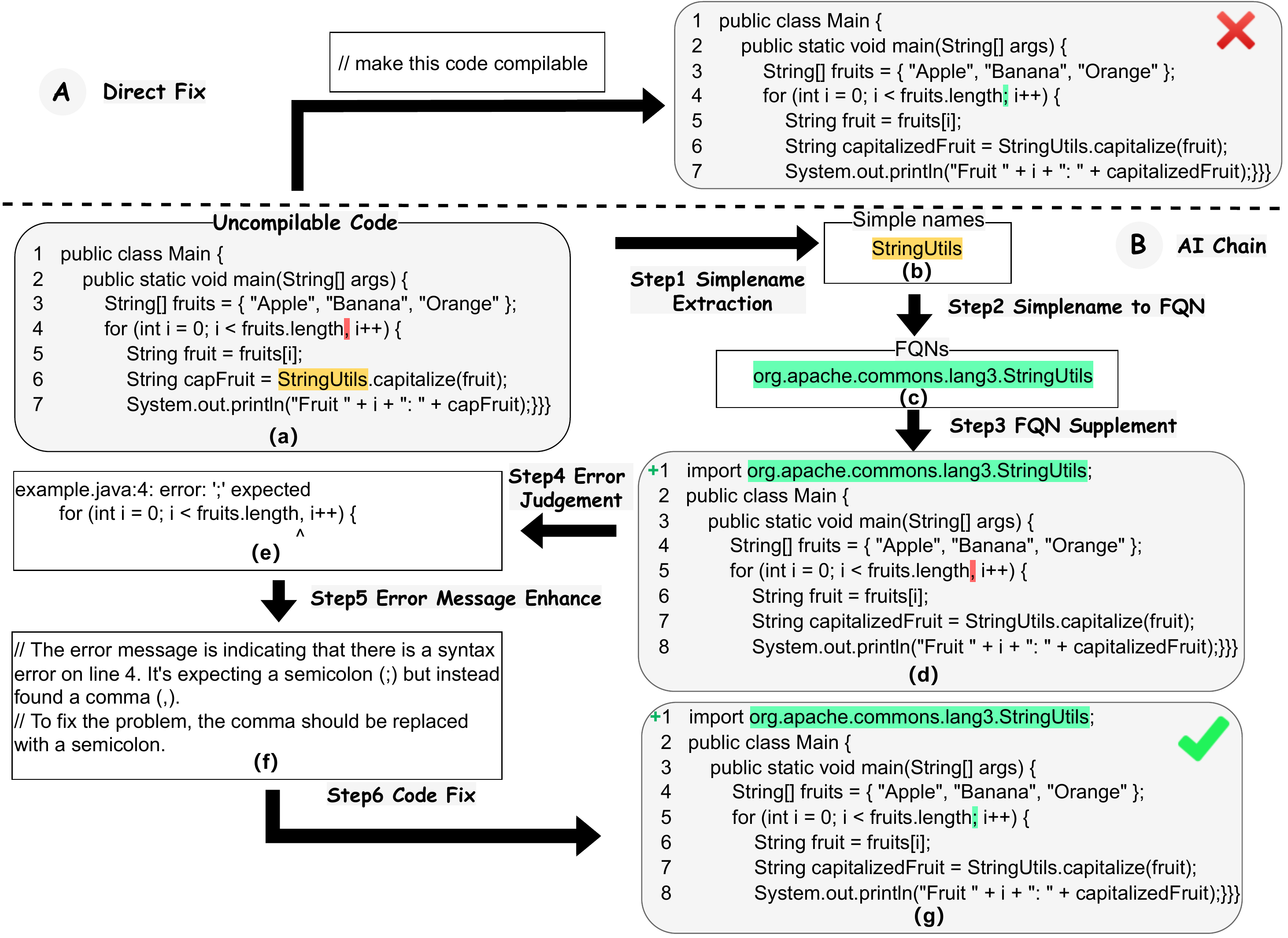}
    \caption{Fix errors directly vs. fix errors with AI Chain}
    \label{fig 1 : killing example}
    \vspace{-6mm}
\end{figure*}

\IEEEPARstart{P}{artial} code from sources like API documentation and Q\&A site is commonly reused in programming tasks \cite{brandt2010example,storey2014r,mao2017survey,ponzanelli2014mining,an2017stack,yang2017stack}.
However, a recent study analyzing 491,906 posts on Stack Overflow found that over 90\% of the partial code is uncompilable~\cite{terragni2016csnippex} due to non-fully qualified names (non-FQNs)~\cite{huang2022prompt,kang2019semantic,saifullah2019learning,huang2022se,Huang2022112PK}, and last-mile syntax errors~\cite{bavishi2022neurosymbolic} such as unbalanced parentheses, missing commas.
This presents a significant challenge for developers, limiting the usefulness of partial code and wasting time and effort. 

The Java code in Fig.~\ref{fig 1 : killing example}-a contains non-fully qualified names (non-FQNs) and last-mile syntax errors that prevent it from compiling correctly. For example, the use of \textit{``StringUtils''} on line 6 is a non-FQN that cannot be resolved by the compiler, resulting in a ``cannot find symbol'' error. In addition, the comma error on line 4 is a last-mile syntax error that causes an ``unexpected token'' error during compilation. These errors collectively prevent the code from running as intended.

When working with partial code, developers typically perform two manual steps to repair it: FQN inference (to infer missing fully qualified names) and syntax error fix (to check and correct last-mile syntax errors). 
However, this process can be time-consuming and error-prone, resulting in inefficiencies during the development process.
To address this issue, many partial program analysis techniques are proposed, including dictionary lookup strategies based on a symbolic knowledge base~\cite{dong2022snr,subramanian2014live} to infer FQNs of non-FQN types, and symbolic-based approaches ~\cite{debroy2010using,rolim2017learning,wang2018search,gulwani2018automated} to fix syntax errors.
However, these methods may encounter out-of-vocabulary (OOV) failures, requiring extensive domain-specific knowledge.


Unlike symbolic-based approaches, which have finite knowledge, recent studies~\cite{Devanbu2012OnTN,Allamanis2018ASO,ray2016naturalness,khanfir2022codebert} propose treating code as text and training large language models (LLMs) on near-infinite code text. 
These LLM-based approaches, such as CodeBERT~\cite{Feng2020CodeBERTAP}, CodeT5~\cite{wang2021codet5}, Copilot~\cite{GithubCopilot}, can effectively reduce out-of-vocabulary (OOV) failures. 
For example, some researchers have used LLM-based approaches to infer FQNs of non-FQN types~\cite{huang2022prompt} and fix last-mile syntax errors~\cite{jiang2021cure}. 
However, these approaches are typically heavyweight supervised learning, requiring large amounts of data and computing resources for efficient model training and gradient updating. 
In contrast, in-context learning (ICL) is a lightweight unsupervised approach proposed by Brown et al.~\cite{brown2020language} that prompts LLMs to learn from a few examples of specific downstream tasks in context and complete those tasks by mimicking those examples~\cite{liu2021makes,min2021metaicl,chan2022data,min2022rethinking}, without gradient updates~\cite{raffel2020exploring, Radford2019LanguageMA, Brown2020LanguageMA}.
Inspired by this, we consider leveraging LLMs' ICL capability to make partial code compileable.

However, directly asking LLMs to make partial code compilable is challenging, as LLMs may only fix some syntax errors but not infer FQNs of non-FQN types. 
Since LLMs are language models, both FQNs and non-FQNs represent correct type names from the point of view of language features, so LLMs may still consider code with non-FQNs to have no compilation errors in the absence of importing packages.
However, this is incorrect. 
Even if code has no syntax errors, it still may not compile because it contains non-FQNs. 
As shown in Fig.~\ref{fig 1 : killing example}-A, when directly querying the LLM with the prompt ``make this code compilable'', the LLM still responds with uncompilable code.

Instead of directly querying LLMs in one step to make partial code compilable, a more effective approach is to design a chain of thought (CoT) with multiple steps~\cite{wu2022ai,Wang2022SelfConsistencyIC,wei2022chain}. 
As shown in Fig.~\ref{fig 1 : killing example}-B, this CoT consists of two main parts: FQN Inference and Syntax Error Fix.
The FQN Inference is further divided into three steps: Simplename Extraction, Simplename to FQN, and FQN Supplement.
These steps comprehensively infer FQNs for simple names in the code.
The Syntax Error Fix is decomposed into three steps: Error Judgement, Error Message Enhance, and Code Fix. 
Together, these steps enhance the error messages and provide solutions for fixing the code, thus resolving syntax errors.
Overall, the CoT includes six key steps that make it an effective solution for making partial code compilable.

However, this CoT still has limitations due to its single prompt implementation, which can lead to error accumulation and an ``epic'' prompt with too many responsibilities. This can lead to errors and difficulties in controlling and improving the prompts.
To overcome these limitations, we adopt the principle of single responsibility in software engineering and decompose the CoT into an AI chain, with each step corresponding to a separate AI unit. We develop an effective prompt for each AI unit, which performs a separate LLM call. As shown in Fig.~\ref{fig 1 : killing example}-B, this AI chain can interact with LLM step by step, thus solving both non-FQN and last-mile syntax error.

We evaluate the effectiveness of our approach, Partial Code Reuse Chain (PCR-Chain), through a series of experiments.
We first test the accuracy of each Unit and Module, with units achieving accuracy rates ranging from 92.1\% to 95.7\%, indicating the effectiveness of prompt design.
We also verify the accuracy of our key modules, FQN Inference and Syntax Error Fix, achieving rates of 80.5\% and 98\%, respectively.

We then compare PCR-Chain to state-of-the-art (SOTA) methods such as BIFI~\cite{yasunaga2021break}, CURE~\cite{jiang2021cure} and RING~\cite{joshi2022repair}, in both dynamically-typed (Python) and statically-typed (Java) programming languages.
In Python, PCR-Chain outperforms SOTA LLM-based methods such as BIFI by 13.7\% accuracy and ICL-based methods such as RING by 5\%.
In Java, our approach accurately solves both non-FQNs and last-mile syntax errors with an accuracy of 80.5\%, while SOTA methods such as CURE and RING could only solve the latter.
We also conduct an ablation experiment to investigate the effectiveness of our AI Chain design principles and find that our design was reasonable. Besides, we explore the sensitivity of our approach to prompt forms and find that our method's accuracy remains stable under different prompt types.

From our study, we derive three AI chain design principles, including hierarchical task decomposition, unit composition, and a combination of AI and non-AI units, that can serve as guidelines for future software engineering projects.

This paper makes the following contributions:
\begin{itemize}


\item{To the best of our knowledge, we are the first to propose the prompt architecture, which combines three global design ideas: hierarchical task breakdown, prompt composition, and a mix of prompt-based AI and non-AI units, rather than a simple AI chain.}
\item{Instead of creating an ``epic'' prompt to implement CoT, we split the CoT into an AI chain, with each step corresponding to a separate AI unit, making the design of the AI chain more reasonable and easier to optimize and control.}
\item{We stand on the giant LLM's shoulder and use a lightweight in-context learning method to resolve FQNs and fix syntax errors in partial code. 
}
\item{The successful completion of the unit, module, and PCR-Chain demonstrates the efficacy of the prompt design, composition, and architecture, to resolve FQNs and fix last-mile syntax errors}.
\end{itemize}

\begin{figure*}
    \centering
    \includegraphics[width=1\textwidth]{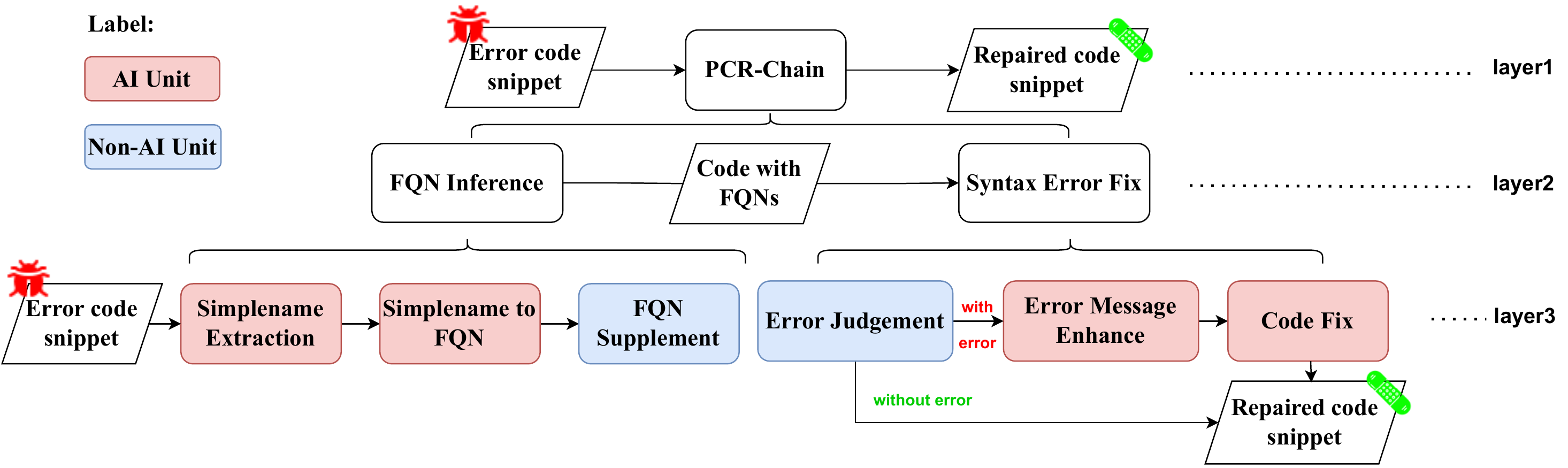}
    \caption{Overall Framework of PCR-Chain}
    \label{fig: Overall framework of PCR-Chain}
\end{figure*}

\begin{figure*}[]
    \centering
    \includegraphics[width=1\textwidth]{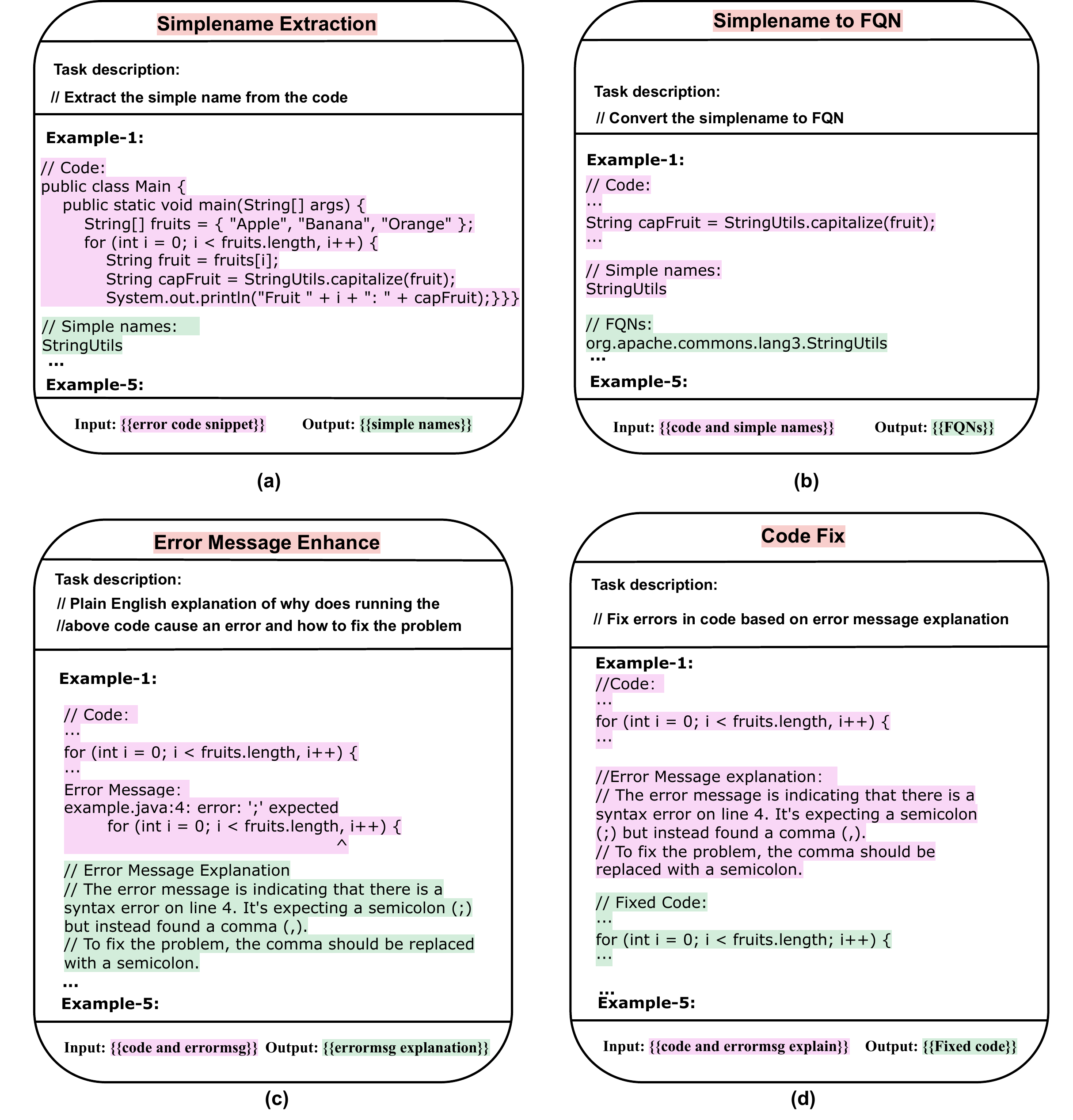}
    \caption{All AI units design in approach}
    \label{fig: all ai units design in approach}
\end{figure*}
\section{APPROACH}
Partial code reuse can be a difficult task due to non-FQNs and last-mile syntax errors. Non-FQNs arise when program entities lack declarations, while last-mile syntax errors occur due to minor mistakes like missing commas or unbalanced parentheses.
To address these challenges, we propose a new approach called the Partial Code Reuse Chain (PCR-Chain). Our approach leverages the vast amount of code knowledge obtained during the training of large language models (LLMs) like ChatGPT\footnote{\href{https://openai.com/blog/introducing-chatgpt-and-whisper-apis}{\textcolor{blue}{https://openai.com/blog/introducing-chatgpt-and-whisper-apis}}}.
PCR-Chain is designed to simulate the human thought process by breaking down the task of making partial code compilable into single-responsibility sub-problems and designing functional units. These units are then connected in a serial or conditional structure to interact with the underlying LLM.
Unlike fine-tuning LLM, which requires significant effort in data gathering, preprocessing, annotation, and model training, PCR-Chain only needs to consider task characteristics, data properties, and information flow, utilizing the capabilities of ChatGPT.

\subsection{Hierarchical Task Breakdown}
As the code contains multiple errors (i.e., non-FQNs and last-mile syntax errors), it is not easy for LLM to resolve all errors in single LLM call.
As shown in Fig.~\ref{fig 1 : killing example}-A, when using a single instruction ``make this code compilable'' to call a single LLM, it only fixes the last-mile syntax error, but not fixes the non-FQN error.
To address this issue, we need to make the instruction more informative and break it down into several sub-instructions, each executed by a separate LLM call, to solve the code error step by step, as shown in Fig.~\ref{fig 1 : killing example}-B.
To facilitate a reasonable decomposition, we re-analyze the code in Fig.~\ref{fig 1 : killing example}(a) and identify that
the issue of partial code reuse can be divided into two sub-modules: the first sub-module is \textbf{FQN Inference}, and the second sub-module is \textbf{Syntax Error Fix} (see Fig.~\ref{fig: Overall framework of PCR-Chain}-layer 2). 

\subsection{Hierarchical Module Decomposition}
In the overall framework of PCR-Chain, as shown in Fig.~\ref{fig: Overall framework of PCR-Chain}-layer 3, the functional units are classified into two categories: AI units that utilize fuzzy reasoning based on LLM, and non-AI units that follow pre-defined rules or logic.
The design of these modules/units adheres to two important software engineering principles, namely separation of concerns and single responsibility, and employs a modular design structure.

Two key modules in the PCR-Chain approach are the \textbf{FQN Inference} module and the \textbf{Syntax Error Fix} module. 

The \textbf{FQN Inference} module is responsible for identifying simple names in the code and inferring them as fully qualified names (FQNs), then completing the FQNs in the code. This module consists of two AI units, namely \textit{Simple Extraction} and \textit{Simple to FQN}, and a non-AI unit called FQN Supplement.
To accomplish this task, the \textbf{FQN Inference} module first extracts simple names from the code using the \textit{Simple Extraction} unit. It then infers the FQNs of the simple names using the \textit{Simple to FQN} unit, then combining the inferred FQNs with the original code using the FQN Supplement unit.

On the other hand, the \textbf{Syntax Error Fix} module aims to enhance error messages and fix syntax errors in code. This module comprises two AI units, \textit{Error Message Enhance} and \textit{Code Fix}, as well as a non-AI unit called Error Judgement. The Error Judgement unit employs a compiler to assess whether there are any errors in the code, and if so, retrieves the error messages for the code. The \textit{Error Message Enhance} unit enhances the error messages and offers solutions for fixing them. Finally, the \textit{Code Fix} unit leverages the enhanced error messages to rectify the syntax errors in the code.

\subsection{Prompt Design for AI-Units}
The AI units in PCR-Chain are designed to activate LLMs for downstream tasks.
There are two methods for implementing this: supervised fine-tuning and in-context learning. 
Supervised fine-tuning involves adjusting the weights of the LLMs using a labeled dataset that is specific to the task.
On the other hand, in-context learning involves conditioning the LLMs on a task description along with some examples of the task, even if it is an unseen one.
While both methods have their own benefits, in-context learning is more straightforward to adopt as it only requires a task instruction along with zero or several examples. In contrast, fine-tuning requires collecting data and making updates to the model, which can be more time-consuming and resource-intensive.

Taking all of these factors into account, our AI units are developed using in-context learning. This approach allows for simpler implementation while still providing effective and efficient performance on a wide range of tasks.

An empirical study~\cite{huang2022se} found that descriptions and examples are critical for in-context learning. 
To standardize our prompt design, we develop a generic template that includes a task description and input-output examples. 
The \textit{Simplename Extraction} unit serves as an example of the template's structure, as shown in Fig.~\ref{fig: all ai units design in approach}(a).
The template includes a task description (e.g., ``Extract the simple names in the code''), followed by five input-output examples (e.g., Input: \textit{``...String capFruit = StringUtils.capitalize(fruit)...''}, Output: \textit{``StringUtils''}). 
After being provided with a code, the LLM extracts the simple names in the code. 

Noted that in this work, we pre-select five examples that are used for all AI units. While the model adaptability generally increases with more examples~\cite{huang2022se}, Min et al.~\cite{min-etal-2022-rethinking} have shown that additional examples beyond four results in limited increase in accuracy.

In the following sections, we describe the prompt designs for each of the four units in the PCR-Chain approach.

\subsubsection{Simplename Extraction Unit}
This AI unit is responsible for extracting the simple names from the given code. 
To prompt the LLM to perform this task, a generic template is used, as shown in Fig.~\ref{fig: all ai units design in approach}(a), with a task description of ``Extract the simple names in the code'', five input-output examples, and a placeholder to input the code to be extracted simple names.
\subsubsection{Simplename to FQN Unit}
This AI unit is responsible for inferring the simple names to FQNs.
To prompt the LLM to perform this task, a generic template is used, as shown in Fig.~\ref{fig: all ai units design in approach}(b), with a task description of ``Convert the simplename to FQN'', five input-output examples, and a placeholder to input the code and simplenames to convert to FQNs.
\subsubsection{Error Message Enhance Unit}
This AI unit is responsible for enhancing the error message from compiler.
To prompt the LLM to perform this task, a generic template is used, as shown in Fig.~\ref{fig: all ai units design in approach}(c), with a task description of ``Plain English explanation of why does running the above code cause an error and how to fix the problem'', five input-output examples, and a placeholder to input the code and corresponding error message.

\subsubsection{Code Fix Unit}
This AI unit is responsible for fixing errors based on error message explanation.
To prompt the LLM to perform this task, a generic template is used, as shown in Fig.~\ref{fig: all ai units design in approach}(d), with a task description of ``Fix errors in code based on error message explanation'', five input-output examples, and a placeholder to input the code and error message explanation.

\subsection{Running Example}
To illustrate how the different units work together and how the data is transformed among them, we present an example using a Java code that contains a non-FQN error and a last-mile syntax error, as shown in Fig.~\ref{fig 1 : killing example}(a).
To start, the Java code is input into the \textit{Simplename Extraction} unit, which identifies the simple names and extracts them. 
The output of this unit is shown in Fig.~\ref{fig 1 : killing example}(b).
Next, the code and simple names are fed into the \textit{Simplename to FQN} unit, which infers the FQNs for the identified simple names. The output of this unit is shown in Fig.~\ref{fig 1 : killing example}(c).
Subsequently, the code and FQNs will be input into the FQN Supplement unit, it will combine code and FQNs.
The output of this unit is shown in fig~\ref{fig 1 : killing example}(d).
Then, the code with FQNs is input into the \textit{Error Judgement} unit, which utilizes a compiler to detect errors and provide error messages. 
The output of this unit is shown in Fig.~\ref{fig 1 : killing example}(e).
After that, the code and corresponding error message are input into the \textit{Error Message Enhance} unit, which provides plain English explanations of why the code produces the error and how to fix the problem. 
The output of this unit is shown in Fig.~\ref{fig 1 : killing example}(f).
Finally, the code and the explanation of the error message are input into the \textit{Code Fix} unit, which fixes the errors and outputs the error-free code, as shown in Fig.~\ref{fig 1 : killing example}(g).

\section{Experiments Setup}
In this section, we present our research questions (RQs) that evaluate the effectiveness of our approach. Additionally, we describe our experimental setup, including data preparation, baselines, and evaluation metrics.

\subsection{Research Questions}\label{RQ}

We conducted the following research questions to evaluate PCR-Chain's performance in partial code reuse.
\begin{itemize}
\item
RQ1: What is the quality of each unit or module in PCR-Chain?
\item
RQ2: How well does PCR-Chain perform in partial code reuse?
\item
RQ3: How effective are the AI Chain and error message enhance strategies employed in PCR-Chain?
\item
RQ4: How sensitive are the prompts in PCR-Chain to different forms?
\end{itemize}

\subsection{Data Preparation}
To evaluate the performance of our approach, we collect two distinct datasets: one for Python and one for Java.
For the Python dataset, we randomly select 200 syntactically invalid code snippets from the dataset used by the state-of-the-art syntax repair tool for Python, BIFI~\cite{yasunaga2021break}. These code snippets were collected from real GitHub repositories.

For the Java dataset, we first crawl 30,000 posts from Stack Overflow that are tagged with ``java'', and collect the highest up-voted answer from each post. Then we extract Java code by identifying the text between the code block HTML tags, i.e., $<$pre$><$code$>$, and identify code with errors using a compiler. We manually filter out 200 error codes that have non-Fully Qualified Names (FQNs) and last-mile syntax errors.

In summary, we prepare two distinct datasets:
\begin{itemize}
    \item Python dataset: 200 error Python code that all contain last-mile syntax issues.
    \item Java dataset: 200 error Java code that all contain non-FQNs and last-mile syntax issues.
\end{itemize}

\subsection{Baselines}
To evaluate the effectiveness of PCR-Chain's overall design and module designs, we compare it with the state-of-the-art methods that reuse partial code. These methods fall into two main approaches: LLM-based and ICL-based.
 

For Python, we compare PCR-Chain with BIFI~\cite{yasunaga2021break}, an LLM-based SOTA method, and RING~\cite{joshi2022repair}, an ICL-based SOTA method. 
In Java, we compare PCR-Chain with CURE, an LLM-based SOTA method, and RING. We modified the examples in RING's prompt to be compatible with Java, making it possible to apply it to Java code.


To conduct the evaluation, we obtain the code for BIFI\footnote{\href{https://github.com/michiyasunaga/BIFI}{\textcolor{blue}{https://github.com/michiyasunaga/BIFI}}} and CURE\footnote{\href{https://github.com/lin-tan/CURE}{\textcolor{blue}{https://github.com/lin-tan/CURE}}} from their respective GitHub repositories, as they are both state-of-the-art methods for reusing partial code in Python and Java, respectively. 
BIFI is implemented using an LLM-based approach in Python, while CURE is implemented in Java using the same approach. 
However, RING does not release its code, so we reproduce it as accurately as possible.

\begin{figure*}
    \centering
    \includegraphics[width=1\textwidth]{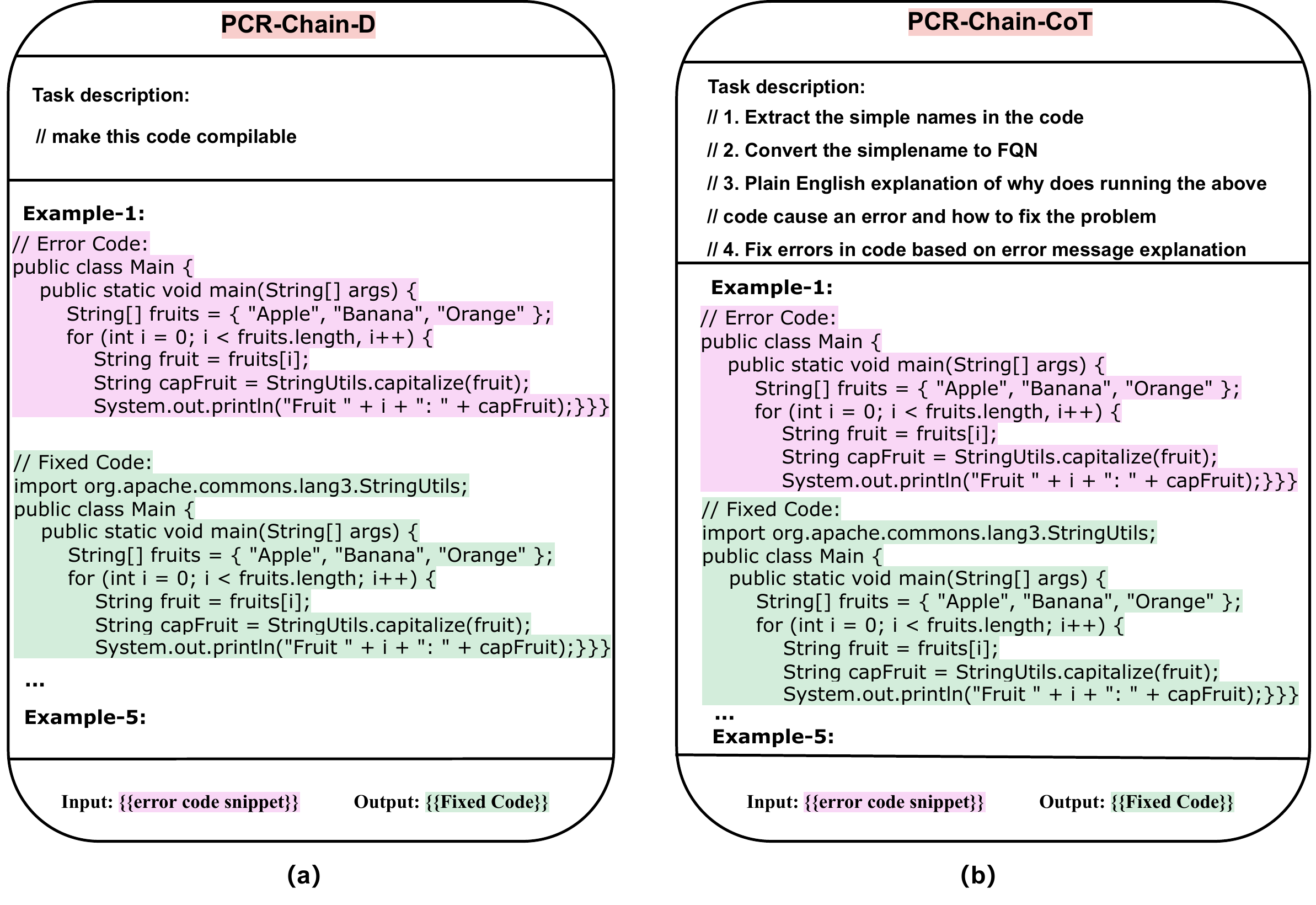}
    \caption{Consult LLM directly (PCR-D) and consult LLM based on CoT (PCR-CoT)}
    \label{fig:D and CoT}
\end{figure*}

To gain a better understanding of the mechanism behind the PCR-Chain, we conduct an ablation study by designing three variants:
\begin{itemize}[leftmargin=*]
    \item
    PCR-D (see Fig.~\ref{fig:D and CoT}(a)), which directly calls the LLM to generate the PCR of the Java code.
    \item
    PCR-CoT (see Fig.~\ref{fig:D and CoT}(b)), a single-prompting approach that describes all steps in one chunk of prompt text and completes a single generative pass.
    \item
    PCR-Chain$_{w/oEME}$, a multiple-prompting approach that does not enhance the error message from compiler, that is without error message enhance (EME).
\end{itemize}

We conduct several experiments to evaluate different aspects of our approach.
First, we compare the effectiveness of two different designs: PCR-D and PCR-CoT.
Next, we compare PCR-CoT with PCR-Chain to evaluate the effectiveness of our AI chain design.
Finally, we evaluate the effectiveness of our error message enhancement (EME) strategy by comparing PCR-Chain$_{w/oEME}$ with PCR-Chain.

\subsection{Evaluation Metrics}\label{metrics}
In our study, we use different metrics to evaluate the effectiveness of our approach for RQs.

For RQ1, we measure the quality of units and modules using the accuracy metric.

For RQ2, RQ3, and RQ4, we use three metrics to evaluate the effectiveness of our approach. These metrics are:
\begin{itemize}
    \item Number of resolved non-FQNs: This metric counts the code snippets that no longer contain non-FQNs after applying our approach.
    \item Number of resolved last-mile syntax errors: This metric counts the code snippets that no longer contain last-mile syntax errors after applying our approach.
    \item Total number of all resolved: This metric counts all the resolved non-FQNs and last-mile syntax errors combined.
\end{itemize}

\section{EXPERIMENTAL RESULTS}

In this section, we explore four research questions to evaluate and discuss the performance of our method.

\subsection{RQ1: What is the quality of each unit or module in PCR-Chain?}

\subsubsection{Motivation}
The CoT approach is a widely adopted method that advocates for breaking down complex tasks into smaller and more manageable steps.
However, relying solely on a single ``epic'' prompt in CoT-based methods can limit their effectiveness and lead to errors accumulating. 
To address this limitation, we develop an AI chain that consists of explicit sub-steps, with each step corresponding to a separate AI unit or non-AI unit.
In this RQ, we investigate whether each AI unit and module in our approach can effectively ensure the accuracy of partial code reuse.

\subsubsection{Methodology}
To evaluate the effectiveness of each AI unit and module in PCR-Chain on Java dataset, we first need to obtain ground-truth data. 
To do so, we enlist the help of two computer science postgraduate students to manually fix unresolved symbols in the Java dataset based on information present in the posts, such as mentioned APIs and API links. One annotator perform the fixing, while the other validate the fixed code. Any disagreements are discussed and resolved.

Once we had the FQNs for non-FQN types in the Java dataset, we could evaluate the \textit{Simplename Extraction} unit and \textit{Simplename to FQN} unit by comparing their outputs with the ground-truth. We consider an exact match to be correct. For the \textit{Error Message Enhance} unit, we do not evaluate the accuracy, as our approach allows this unit to be noisy. 
As long as some relevant and correct error message explanation is inferred, the subsequent \textit{Code Fix} unit could make repairs.

Regarding the \textit{Code Fix} unit, we consider an output as correct if the code snippet is free of any last-mile syntax errors.
The results are presented in Table~\ref{RQ1:The Quality of AI Units and Modules}, and more detailed information can be found in Section~\ref{metrics}.

\subsubsection{Result Analysis}
The second column of Table~\ref{RQ1:The Quality of AI Units and Modules} presents the accuracy of each unit. Specifically, the \textit{Simplename Extraction} unit of the \textbf{FQN Inference Module} correctly predicts 95.7\% (815 out of 851) of the simple names in the 200 code snippets, while the \textit{Simplename to FQN} unit infers 92.1\% (781 out of 851) of the corresponding FQNs.
In contrast, for the \textbf{Syntax Error Fix Module}, the \textit{Code Fix} unit accurately fixes 98\% (196 out of 200) of the code snippets that contain last-mile syntax errors.

The accuracy of each module is shown in the fourth column of Table~\ref{RQ1:The Quality of AI Units and Modules}. For the \textbf{FQN Inference} module, it takes in 200 code snippets and outputs 161 code snippets without any non-FQNs, resulting in an accuracy of 80.5\%. On the other hand, for the \textbf{Syntax Error Fix} module, it takes in 200 code snippets and outputs 196 code snippets that are free of last-mile syntax errors, achieving an accuracy of 98\%.

\vspace{1.5mm}
\noindent\fbox{
\begin{minipage}{8.6cm} 
\textit{The high accuracy of the units confirms the usefulness of the prompt design and lays the foundation for high-quality modules. 
The successful execution of modules shows the value of prompt composition in linking units to achieve superior outcomes for higher-level tasks.}
\end{minipage}}

\begin{table}[]
\caption{The Quality of AI Units and Modules}
\label{RQ1:The Quality of AI Units and Modules}
\centering
\begin{tabular}{|c|c|c|l|}
\hline
AI unit               & Acc & Module                            & Acc               \\ \hline
Simplename Extraction & 0.957& \multirow{2}{*}{FQN Inference}    & \multirow{2}{*}{0.805} \\ \cline{1-2}
Simplename to FQN     & 0.921&                                   &                   \\ \hline
Code Fix              & 0.980& \multicolumn{1}{l|}{Syntax Error} &0.980               \\ \hline
\end{tabular}
\end{table}

\subsection{RQ2: How well does PCR-Chain perform in partial code reuse?}
\subsubsection{Motivation}
Our objective is to evaluate the effectiveness of our method in terms of partial code reuse and compare it with the SOTA methods such as BIFI~\cite{yasunaga2021break}, CURE~\cite{jiang2021cure} and RING~\cite{joshi2022repair} in both dynamically-typed (Python) and statically-typed (Java) programming languages.

For Python, we will compare our method with two SOTA methods, RING (ICL-based) and BIFI (LLM-based). 
Similarly, for Java, we will compare our method with two SOTA methods, RING (ICL-based) and CURE (LLM-based).

To evaluate the performance of these methods, we will use three metrics: the number of resolved non-FQNs, the number of resolved last-mile syntax errors, and the total number of all resolved, which refers to all non-FQNs and last-mile syntax errors that are resolved.

\subsubsection{Methodology}
We evaluate the performance of PCR-Chain, BIFI, and RING on the Python dataset, and PCR-Chain, CURE, and RING on the Java dataset.
The results are presented in Table~\ref{table: RQ2 result} and more detailed information can be found in Section~\ref{metrics}.

\subsubsection{Result Analysis}
Our evaluation demonstrates that our method is effective in addressing partial code reuse in both dynamically-typed (Python) and statically-typed (Java) programming languages.

For Python, we fix the last-mile syntax errors included in the Code.
As shown in Table~\ref{table: RQ2 result}, our method outperforms all baselines, with an accuracy of 99\% in resolving 198 last-mile syntax errors. 
Notably, our accuracy is about 5\% and 13.7\% higher than RING and BIFI, respectively.
This demonstrates that there is sufficient syntax knowledge in LLM to efficiently address last-mile syntax errors.

For Java, we address both non-FQNs and last-mile syntax errors simultaneously, achieving an accuracy of 80.5\%.
As shown in Table ~\ref{table: RQ2 result}, compared with RING and CURE, our method can effectively solve both non-FQNs and last-mile syntax errors, while RING and CURE only address the latter. 
This difference in performance can be attributed to the fact that RING and CURE directly ask LLMs to make partial code compilable, which can be a challenging task. 
However, LLMs may only fix some syntax errors and struggle to infer FQNs for non-FQN types.

Since LLMs are language models, both FQNs and non-FQNs represent correct type names from the point of view of language features.
This can lead LLMs to mistakenly regard code with non-FQNs as compilable, even if it contains errors due to non-imported packages.
However, in reality, such code may fail to compile despite having no syntax errors, precisely because it includes non-FQNs.

As shown in Fig.~\ref{fig 1 : killing example}-A, when directly querying the LLM with the prompt ``make this code compilable'', the LLM still responds with uncompilable code.
In contrast, our method interacts with LLMs step-by-step using the AI chain, thus allowing us to effectively address both non-FQNs and last-mile syntax errors.
\noindent\fbox{
\begin{minipage}{8.6cm} 
\textit{Standing on the shoulder of LLM for partial code reuse, PCR-Chain effectively resolves FQNs and fixes last-mile syntax errors.
Each AI unit in the AI Chain follows the principle of single responsibility and can interact with LLMs separately to effectively reuse partial code.}
\end{minipage}}

\begin{table*}[h]
\centering
\caption{Evaluations Results of Multiple Languages (``-'' means this error do not contain in this dataset)}
\label{table: RQ2 result}
\begin{tabular}{|c|c|c|c|c|}
\hline
Dataset                 & Approach & Resolved Non-FQNs & Resolved Last-mile Syntax Errors & All Resolved \\ \hline
\multirow{3}{*}{Python} & RING     & -                 & 188                              & 188          \\ \cline{2-5} 
                        & BIFI     & -                 & 174                              & 174          \\ \cline{2-5} 
                        & PCR      & -                 & \textbf{198}                              & \textbf{198}          \\ \hline
\multirow{3}{*}{Java}   & RING     & 0                 & 188                              & 0            \\ \cline{2-5} 
                        & CURE     & 0                 & 137                              & 0            \\ \cline{2-5} 
                        & PCR      & \textbf{161}               & \textbf{196}                              & \textbf{161}          \\ \hline
\end{tabular}
\end{table*}

\subsection{RQ3: How effective are the AI Chain and error message enhance strategies employed in PCR-Chain?}
\subsubsection{Motivation}
CoT can alleviate the challenge posed by directly relying on LLMs, such as the challenge of LLMs mistakenly regarding code with non-FQNs as compilable. 
However, CoT-based approaches face difficulties in control and optimization due to the ``epic'' cues with excessive accountability. 
To solve this problem, we design an AI chain that facilitates step-by-step interaction with the LLM to effectively tackle both non-FQN and last-mile syntax errors. 

In this RQ, we aim to investigate two aspects of our approach. Firstly, we would like to explore whether our AI chain design can effectively interact with LLMs, thus enhancing the robustness of our approach. Secondly, we would like to investigate whether the error message enhance strategy could enhance the effectiveness of our AI chain.

\subsubsection{Methodology}
We set up three approach variants (PCR-D, PCR-CoT, and PCR-Chain$_{w/oEME}$). 
Two scenarios (PCR-D vs. PCR-CoT, PCR-CoT vs. PCR-Chain) are used to evaluate the effectiveness of the AI chain.
The last one scenario (PCR-Chain$_{w/oEME}$ vs. PCR-Chain) is used to evaluate the effectiveness of error message enhance strategy.
The results are presented in Table~\ref{table: RQ3 Results}, and more detailed information can be found in Section~\ref{metrics}.

\subsubsection{Result Analysis}

\begin{table*}[]
\centering
\caption{Ablation Results of PCR-Chain Variants (``-'' means this error do not contain in this dataset)}
\label{table: RQ3 Results}
\begin{tabular}{|c|c|c|c|c|}
\hline
Strategies                & Dataset & Resolved Non-FQNs & Resolved Last-mile Syntax Errors & All Resolved \\ \hline
\multirow{2}{*}{PCR-Chain}& Java    &161                &196                               &161           \\ \cline{2-5} 
                          & Python  &-                  &198                               &198           \\ \hline
\multirow{2}{*}{PCR-D}    & Java    &101                &188                               &101           \\ \cline{2-5} 
                          & Python  &-                  &188                               &188           \\ \hline
\multirow{2}{*}{PCR-CoT}  & Java    &141                &190                               &141           \\ \cline{2-5} 
                          & Python  &-                  &190                               &190           \\ \hline
\multirow{2}{*}{\begin{tabular}[c]{@{}c@{}}PCR-Chain$_{w/oEME}$\end{tabular}} & Java    &146&190&146   \\ \cline{2-5} 
                          & Python  &-                  &180                               &180          \\ \hline
\end{tabular}
\end{table*}

The experimental results are presented in Table~\ref{table: RQ3 Results}.
For PCR-D, both the number of resolved non-FQNs and resolved last-mile syntax errors are fewer than PCR-CoT.
This is because when directly asking LLMs to make code compilable, LLMs may mistakenly regard code with non-FQNs as compilable, even if it has missing imported packages.
As shown in Fig.~\ref{fig 1 : killing example}-A, when directly querying the LLM with the prompt ``make this code compilable'', the LLM still responds with uncompilable code.
However, the CoT-based prompt is more informative than PCR-D's prompt. 

PCR-CoT shows fewer number of resolved non-FQNs and resolved last-mile syntax errors compared to PCR-Chain. 
This highlights the superiority of our AI chain design over CoT's single-prompting approach, which uses an ``epic'' prompt with hard-to-control behavior and error accumulation.

In contrast, PCR-Chain breaks down the CoT into separate AI units, allowing step-by-step interaction with LLMs for partial code reuse. The last two rows of the Table~\ref{table: RQ3 Results} demonstrate that the error message enhancement strategy improves PCR-Chain's effectiveness in reusing partial code.

\noindent\fbox{\begin{minipage}{8.6cm} \emph{Compared with directly asking the LLM to achieve code reuse, our AI chain design that interacts with the LLM can effectively improve the response reliability of the LLM.
Our error message enhancement strategy can reuse partial code more effectively.}\end{minipage}}

\subsection{RQ4: How sensitive are the prompts in PCR-Chain to different forms?}

\subsubsection{Motivation}
To further explore the impact of prompt forms on results, we investigate the task description, demonstration examples, and content representation of the prompt.

\subsubsection{Methodology}
To investigate the impact of task description, we explore prompts with and without task descriptions. Additionally, we examine the impact of the order of demonstration examples, considering different orders such as order, reverse order, and fixed order based on cosine similarity between the prompt's input and the examples.

For content representation, we focus on the format of representation, whether it is in natural language or a semi-structured form using specific tags.
For example, the prompt can be expressed as either \textit{``Task Description: Extract the simple name from the code.''} in natural language form or as \textit{``$<$Task Description$>$ Extract the simple name from the code$<$/Task Description$>$.''} in a semi-structured form.
We use the Java dataset in this RQ.

Before conducting our experiments, we define a basic configuration based on our intuition of what options would be most effective for PCR-Chain.
The basic configuration includes a task description, example prompts in a fixed order, and the use of natural language form in the prompt.
We then create variant configurations by changing one factor at a time (task description, order of demonstration examples, content representation), while keeping the other two factors the same as in the basic configuration.

\subsubsection{Result Analysis}
The results reveal that including a task description improves the LLM's understanding of tasks, as there were two fewer effective code reuses observed without a task description compared to with a task description.

The order of demonstration examples significantly impacts the LLM's effectiveness. 
Presenting the most similar example first (i.e., with the most similar example farthest from the prompt's input) resulted in the worst accuracy, with 153 out of 200 codes reused, while presenting the most dissimilar example first (i.e., with the most similar example closest to the prompt's input) led to the best accuracy, with 162 out of 200 codes reused. This indicates that the LLM is highly influenced by the example closest to the prompt's input.The results of content representation experiment shows that using natural language prompts reuses 8 more code snippets than using semi-structured forms. This highlights the importance of natural language over semi-structured forms in promoting effective code reuse. While semi-structured forms may improve prompt organization and clarity, they weaken the LLM's learning ability. In contrast, natural language prompts facilitate better learning and enhance the LLM's effectiveness.
\begin{table*}[]
\caption{Results of Sensitivity of Prompt (+/-Value Against The Basic Config)}
\centering
\label{table: RQ4 Results}
\begin{tabular}{|c|c|c|c|c|}
\hline
Factor                                           & Variant          & Resolved Non-FQNs & \begin{tabular}[c]{@{}c@{}}Resolved Last-mile\\  Syntax Errors\end{tabular} & All Resolved \\ \hline
                                                 & Basic Config     &161                &196                                                                          &161         \\ \hline
Task Description                                 & Not Provided     &\textcolor{red!90!black}{-2} &\textcolor{red!90!black}{-3}&\textcolor{red!90!black}{-2}\\ \hline
\multirow{2}{*}{Order of Demonstration Examples} & Similar First &\textcolor{red!90!black}{-8}&\textcolor{red!90!black}{-4}&\textcolor{red!90!black}{-8}\\ \cline{2-5} 
                                                 & Dissimilar First    &\textcolor{green!70!black}{+1}&\textcolor{green!70!black}{+1}&\textcolor{green!70!black}{+1}\\ \hline
Content Representation                           & Semi-Structured         &\textcolor{red!90!black}{-8}&\textcolor{red!90!black}{-2}&\textcolor{red!90!black}{-8}\\ \hline
\end{tabular}
\end{table*}
\vspace{2mm}
\noindent\fbox{\begin{minipage}{8.6cm} \emph{ 
Although studies~\cite{gao2020making,ding2021prompt,zhao2021calibrate} show LLMs are sensitive to prompt factors, our approach remains overall stable in reusing partial code in face of variant prompt factors. Our intuition of the effectiveness of factor variants largely holds, except for prompt order of demonstration example. This indicates the necessity to combine intuition and empirical evidences in prompt design.
}\end{minipage}}

\section{DISCUSSION}
In this section, we summarize the principles of AI chain and prompt design patterns, and also discuss potential threats to validity. 

\subsection{Prompt Engineering Principles}
Our experiments highlight the importance of improving the reliability of LLM responses through the design of an informative CoT and the creation of an effective AI chain with multiple single-responsibility, composable steps. 

In designing an AI chain, we propose three principles:
\begin{itemize}
    \item Hierarchical Task Breakdown: Breaking down a problem into modules, submodules, and further breaking them down into functional units, facilitating a structured problem-solving approach.
    \item Unit Composition: Connecting functional units in a specific structure enables cohesive functioning of the AI chain and achieves optimal results.
    \item Mixing of AI Units and non-AI Units: Designing logic functional units as non-AI units, and utilizing the LLM for fuzzy logic functional units through prompt design.
\end{itemize}

We believe that prompt engineering will play a crucial role in the future of problem-solving. 
The principles  offer valuable guidance for designing AI chains and maximizing the potential of LLM-based paradigms for effective problem-solving.

\subsection{Threats to Validity}
Our method faces both internal and external threats.
An internal threat is the potential inconsistency in the manually labeled ground-truth data in the Java dataset.
To mitigate this, we employ two annotators to label simultaneously and measure the consistency of the results using the Kappa coefficient. A high Kappa coefficient (all coefficients above 80\%) indicates the reliability of the annotation results.

In terms of external threats, our research on partial code reuse has focused on the Java and Python. However, to expand the universality of our approach, we plan to investigate code reuse in niche languages such as smart contract code.

Compared to traditional code reuse methods that require expertise in program analysis and significant engineering and maintenance efforts for different languages and their versions, our method offers greater adaptability. 
Adapting our approach to other languages mainly involves substituting the language type in the prompt examples.

The emergence of new large-scale LLMs, such as GPT-4~\cite{nori2023capabilities, lyu2023translating}, may impact our method. Although we are currently on the waitlist for access to GPT-4, we eagerly anticipate using it in the future to validate the effectiveness and universality of our approach.

\section{Related Work}
Partial code reuse is a common practice among developers, involving the copying and pasting of code snippets from online resources like Stack Overflow into Integrated Development Environments (IDEs). 
However, unresolved type and last-mile syntax errors often hinder the compilation of partial code.
To address these issues, several related works have been conducted in recent years.

\begin{table}[h]
\centering
\caption{Method Comparison}
\label{table:method comparison}
\resizebox{0.5\textwidth}{!}{
\begin{tabular}{|c|c|c|c|l}

\cline{1-4}
          & Non-FQN & \begin{tabular}[c]{@{}c@{}}Last-mile\\ Syntax Error\end{tabular} & Approach-Based &  \\ \cline{1-4}
SNR~\cite{dong2022snr}      &\CheckmarkBold         &\XSolidBrush                        & Symbolic-Based      &  \\ \cline{1-4}
Sumit~\cite{gulwani2018automated}      &\XSolidBrush         &\CheckmarkBold                        & Symbolic-Based      &  \\ \cline{1-4}
Huang~\cite{huang2022prompt}      &\CheckmarkBold         &\XSolidBrush                        & LLM-Based      &  \\ \cline{1-4}
CURE~\cite{jiang2021cure}      &\XSolidBrush         &\CheckmarkBold                        & LLM-Based      &  \\ \cline{1-4}
RING~\cite{joshi2022repair}      &\XSolidBrush         &\CheckmarkBold                        & ICL-Based      &  \\ \cline{1-4}
PCR-Chain &\CheckmarkBold         &\CheckmarkBold                        & CoT-Based      &  \\ \cline{1-4}
\end{tabular}
}
\end{table}

Table~\ref{table:method comparison} presents a comparison of different methods for resolving non-fully qualified name (non-FQN) errors and fixing last-mile syntax errors.

Symbolic-based methods like SNR~\cite{dong2022snr} and Sumit~\cite{gulwani2018automated} are commonly used to resolve non-fully qualified name (non-FQN) errors or fix last-mile syntax errors. However, these methods may encounter out-of-vocabulary (OOV) failures and require significant engineering effort to develop and program analysis/fix experience.

Built on source code naturalness, recent approaches have overcome these by training or fine-tuning a LLM.
For example, Huang et al.~\cite{huang2022prompt}utilize the prompt-tuned LLM to resolve types, and CURE~\cite{jiang2021cure} use a LLM fine-tuned by an automated program repair task to fix syntax errors. 
However, LLM-based methods typically require a large amount of data and computing resources to train efficient models.
Instead, Brown et al.~\cite{brown2020language} propose a more lightweight approach called in-context learning (ICL).

ICL is a novel paradigm that allows foundation models to adapt to new tasks without extensive training or updates. 
Instead of relying on gradient updates~\cite{raffel2020exploring, Radford2019LanguageMA, Brown2020LanguageMA}, ICL utilizes zero- or few-shot prompts for task adaptation.
This paradigm has been successfully applied in range of software engineering tasks, such as testing~\cite{Chen2022CodeTCG}, code generation~\cite{Mastropaolo2023OnTR}, and GUI automation~\cite{Liu2022FillIT}.
RING\cite{joshi2022repair} used an ICL-based method to fix last-mile syntax errors, avoiding the need to train or fine-tune a LLM.
However, this approach does not address non-FQN errors.

Our method, PCR-Chain, differs from RING by not directly querying the LLM with simple prompts like ``make this code compilable.''
Instead, we leverage the idea of CoT to address both non-FQN and last-mile syntax errors. However, existing CoT approaches only provide simple instructions like ``let's do something step by step.'' which cannot handle complex tasks. In contrast, our approach is based on an AI chain~\cite{wu2022ai, wu2022promptchainer,dang2022prompt} that interacts with the model to reuse partial code. 
While the idea of an AI chain has been explored for writing assistants~\cite{wu2022ai}, our AI chain involves complex task analysis and data flow for domain-specific partial code reuse.

\section{CONCLUSION AND FUTURE WORK}
In this paper, we propose a novel approach for partial code reuse that utilizes the vast amount of code knowledge stored in LLMs.
Our approach involves a CoT consisting of six steps, including Simplename Extraction, Simplename to FQN, FQN Supplement, Error Judgement, Error Message Enhance, and Code Fix.

To adhere to the single responsibility principle, we decompose the CoT into an AI chain and supplement it with effective prompt instructions. Our approach outperforms traditional LLM-based methods and the original CoT method in terms of code reuse rates. 
With the lower cost of building LLM-based partial code reuse tools compared to traditional symbolic-based approaches, our method offers a new LLM-based alternative for software engineering tool development.

Our approach provides a practical and cost-effective solution for software engineering tools, reducing the need for extensive engineering and maintenance work.
We also introduce practical principles for using just-in-time engineering and AI chains in SE tasks, demonstrating the potential of LLM4SE. By leveraging the underlying model, we can focus on identifying the problems that AI needs to solve, rather than investing in data collection, labeling, model training, or program analysis. 
Our code and data package can be found here.\footnote{\href{https://github.com/SE-qinghuang/A-Chain-of-AI-based-Solutions-for-Resolving-FQNs-and-Fixing-Syntax-Errors-in-Partial-Code}{\textcolor{blue}{https://github.com/SE-qinghuang/A-Chain-of-AI-based-Solutions-for-Resolving-FQNs-and-Fixing-Syntax-Errors-in-Partial-Code}}}

\section{Acknowledgements}
The work is supported by National Nature Science Foundation of China (Nos. 62262031), and Science and Technology Key Project of Education Department of Jiangxi Province GJJ2200302, GJJ2200303, GJJ2200304, GJJ210307).

\balance
\bibliography{sample-base}

\par\noindent 
\parbox[t]{\linewidth}{
\noindent\parpic{\includegraphics[height=3.0in,width=1in,clip,keepaspectratio]{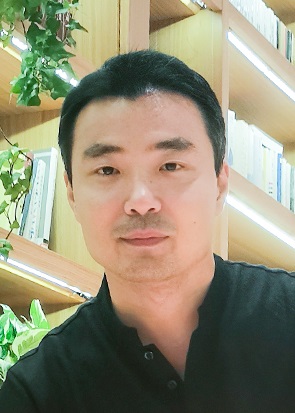}}
\noindent {\bf QING HUANG}\
is an Associate Professor in the School of Computer and Information Engineering, Jiangxi Normal University. His research interests are software engineering and knowledge graph.}
\vspace{0.2\baselineskip}

\par\noindent 
\parbox[t]{\linewidth}{
\noindent\parpic{\includegraphics[height=3.0in,width=1in,clip,keepaspectratio]{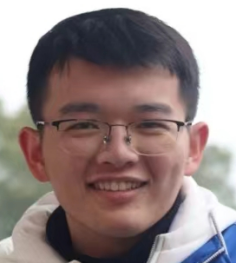}}
\noindent {\bf Jiahui Zhu}\
is a first-year graduate student in the School of Computer and Information Engineering, Jiangxi Normal University. His research interests are software engineering and program repair.
}
\vspace{1\baselineskip}

\par\noindent 
\parbox[t]{\linewidth}{
\noindent\parpic{\includegraphics[height=3.0in,width=1in,clip,keepaspectratio]{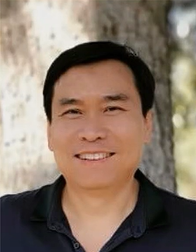}}
\noindent {\bf Zhenchang Xing}\
is an Associate Professor in the Research School of Computer Science, Australian National University. His research areas are software engineering, applied data analytics, and human-computer interaction.}
\vspace{1\baselineskip}

\par\noindent 
\parbox[t]{\linewidth}{
\noindent\parpic{\includegraphics[height=3.0in,width=1in,clip,keepaspectratio]{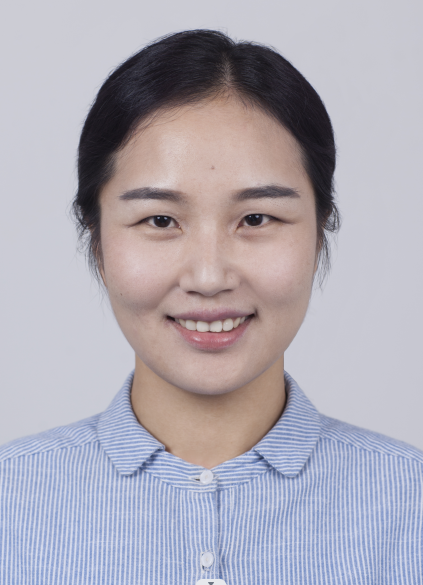}}
\noindent {\bf Huan Jin}\
is currently an associate professor at Jiangxi University of Technology's School of Information Engineering. Her research interests are in data mining and service-oriented software engineering.}
\vspace{1\baselineskip}

\par\noindent 
\parbox[t]{\linewidth}{
\noindent\parpic{\includegraphics[height=3.0in,width=1in,clip,keepaspectratio]{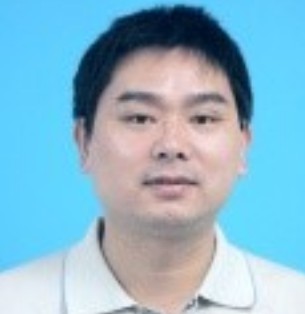}}
\noindent {\bf Changjing Wang}\
is a Professor in the School of Computer and Information Engineering, Jiangxi Normal University, China. His research interests are Web service and formal method.}
\vspace{1\baselineskip}

\par\noindent 
\parbox[t]{\linewidth}{
\noindent\parpic{\includegraphics[height=3.0in,width=1in,clip,keepaspectratio]{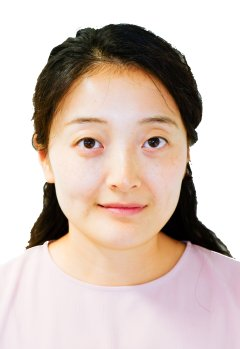}}
\noindent {\bf Xiwei Xu}\
is a Senior Research Scientist with Architecture\& Analytics Platforms Team, Data61, CSIRO. She is also a Conjoint Lecturer with UNSW. She started working on blockchain since 2015. Her main research interest is software architecture. She also does research in the areas of service computing, business process, and cloud computing and dependability.}

\end{document}